# Irradiance Variations During This Solar Cycle Minimum

Thomas N. Woods

*Laboratory for Atmospheric and Space Physics, University of Colorado, 1234 Innovation Drive, Boulder, CO 80303, USA*

**Abstract.** The current cycle minimum appears to be deeper and broader than recent cycle minima, and this minimum appears similar to the minima in the early 1900s. With the best-ever solar irradiance measurements from several different satellite instruments, this minimum offers a unique opportunity to advance our understanding of secular (long-term) changes in the solar irradiance. Comparisons of the 2007-2009 irradiance results to the irradiance levels during the previous minimum in 1996 suggest that the solar irradiance is lower in this current minimum. For example, the total solar irradiance (TSI) indicates lower irradiance in 2008 than in 1996 by about 200 ppm, and the SOHO Solar EUV Monitor (SEM) 26 to 34 nm irradiance is about 15% lower in 2008 than in 1996. However, these irradiance variations have 30-50% uncertainties due to these results depending strongly on instrument degradation trends over 12 years. Supporting evidence for lower irradiance include that the solar magnetic fields are lower and that there are more low-latitude coronal holes during this current minimum.

## 1. Introduction

There are expectations that the solar irradiance could be different from one solar cycle minimum to another based on solar-climate studies, such as during the colder period in the late 1600s that is associated with low solar activity called the Maunder Minimum (e.g., Eddy 1976). The solar cycle minima during the space-age, but prior to this minimum in 2007-2009, appear to have similar irradiance levels based on total solar irradiance (TSI) and solar ultraviolet (UV) irradiance measurements made continuously on satellites from 1978 to the present. This current cycle minimum, between solar cycle 23 and 24, is clearly different than the last few minima in that open magnetic flux is about 30-40% less, solar wind pressure is about 40% lower, and there are more low-latitude coronal holes, which in turn provide more high-speed solar wind streams towards Earth. With lower magnetic activity, there is expected to be a reduction in TSI. With more low-latitude coronal holes, there is expected to be reduced solar coronal emissions that dominate in the extreme ultraviolet (EUV) range. While there are indications that there is lower solar irradiance in 2008 relative to 1996 (previous minimum for cycle 22/23), there is concern that the secular (long-term) trend in the irradiance time series has high uncertainty because of relatively large corrections for instrument degradation between 1996 and 2008. An overview of the comparison of the sunspot number (SSN) record over the past three centuries is given first to put this cycle minimum in perspective of the historical solar record. Then comparisons of the irradiance in 2008 to 1996 are discussed in the remaining sections.

The monthly-averaged SSN time series and comparison of this cycle minimum to the other minima are shown in Figure 1. The characteristics displayed for the cycle minima are the SSN level, solar cycle period, and duration of each cycle minimum. It is striking that it is necessary to go back nine solar cycles, to the early 1900s, to find a cycle minimum that is similar to this current cycle minimum. The cycle minima that appear similar to this current minimum in 2007-2009 are identified with vertical dashed lines and labeled as the Dalton Minimum period in the early 1800s and as the 1900s Minima. The minima in 1798 (cycle 4/5) and in 1879 (cycle 11/12) appear the most similar to this current minimum (cycle 23/24). These three cycle minima occur after a modestly large cycle maximum that is followed by a lower and wider cycle minimum.

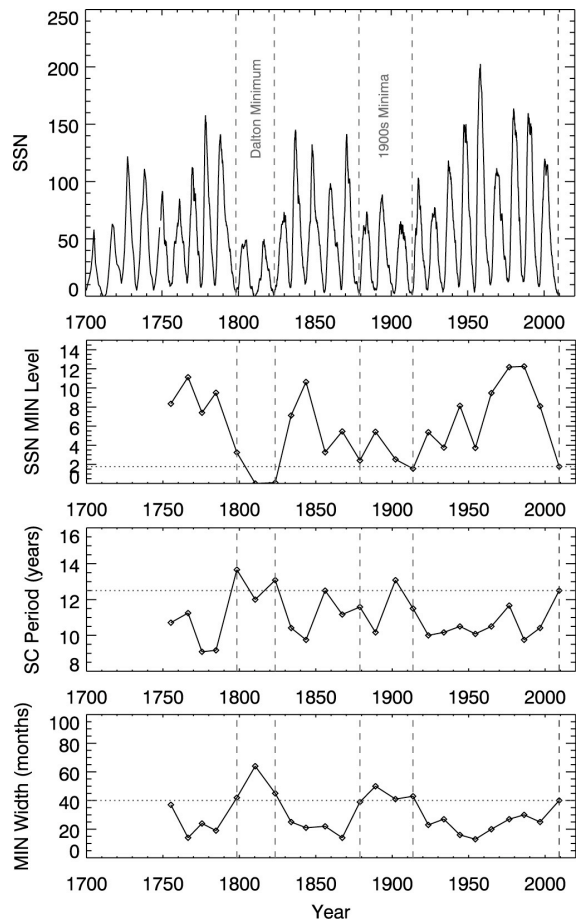

Figure 1. Comparison of Minima for the Sunspot Number (SSN). The SSN record for this current cycle minimum, whose results are indicated as horizontal dotted lines in the bottom three panels, indicates lower SSN minimum, wider solar cycle period, and wider cycle minimum period than the more recent cycle minima. This cycle minimum is more comparable to the cycle minima during the early 1900s and even similar to those during the Dalton Minimum in the early 1800s; both of these times are marked as vertical dashed lines.

## 2. Total Solar Irradiance Variation

There are four different sets of TSI radiometers that have measured the solar irradiance during the cycle 23/24 minimum. These current measurements are combined with previous TSI measurements to make composite time series since 1978. With a specific radiometer being the reference, there are three composite TSI records. These composites are referred to as the Physikalisch-Meteorologisches Observatorium Davos (PMOD, Fröhlich 2006, Fröhlich 2009), Royal Meteorological Institute of Belgium (RMIB, Dewitte et al. 2005, Mekaoui & Dewitte 2008), and Active Cavity Radiometer Irradiance Monitor (ACRIM, Willson & Mordvinov 2003, Scafetta & Willson 2009) composites. The fourth TSI data set, not yet included in its own composite, is from the Total Irradiance Monitor (TIM) aboard the NASA Solar Radiation and Climate Experiment (SORCE) satellite (Kopp, Lawrence, & Rottman 2005). While there are larger than expected differences in the absolute TSI levels from these different instruments (e.g., Kopp et al. 2005), their day-to-day variations agree very well. And of those, the two SOHO-based composite sets (PMOD and RMIB) have been measuring the TSI since the previous minimum in 1996.

Of particular interest, is the TSI level different during this extended minimum (2007-2009) as compared to the 1996 cycle minimum? The four TSI time series, as shown in Figure 2, offer conflicting answers to this question. The PMOD and ACRIM composite time series indicate that this current cycle minimum TSI is about 200 ppm less than in 1996; whereas, the RMIB composite time series suggest that this current TSI level is now higher by about 50 ppm than in 1996. It is interesting to note that the TSI data sets are in best agreement after 2000.

The uncertainty for the PMOD and RMIB irradiance variation between 1996 and 2008 is estimated to be about 100 ppm (Fröhlich 2009, Mekaoui & Dewitte 2008), and this uncertainty may also be appropriate for the ACRIM results. Therefore, the TSI 200 ppm decrease between 1996 and 2008 only has a signal-to-noise ratio of two at best. The TSI composite differences between 1996 and 2000 is as much as 200 ppm, larger than the 100 ppm estimated uncertainty for the TSI variation, so there may be even more uncertainty in tracking instrument degradation and/or in combining different TSI measurements to make composite time series.

Improved understanding of the TSI results at this cycle minimum is of utmost importance for confirming the established solar-climate relationships dating back to before the pre-industrial period; that is, before anthropogenic greenhouse gases influenced climate change. There are several reconstructed TSI records for the Maunder Minimum (1650-1715) and Dalton Minimum (1798-1820) periods when lower solar activity is thought to have caused cooler global temperatures for Earth. The more recent reconstructed TSI records estimate that the cycle-averaged TSI during Maunder Minimum is about 1 W/m$^2$ (700 ppm) less than in 1996 (e.g., Wang, Lean, & Sheeley 2005, Tapping et al. 2007, Steinhilber, Beer, & Fröhlich 2009). These estimates use the solar magnetic flux to determine the secular (long-term) changes in the TSI. Other TSI models that only use sunspot area proxies and chromospheric proxies for the faculae can not predict long-term trends because these proxies indicate the same low level at every minimum.

Because the solar magnetic field has been declining since about 1985, the long-term TSI trend is also expected to be declining. Lockwood, Owens, & Rouillard (2009) show that the open magnetic field flux has decreased by about a factor of two since 1985, which includes about 30% reduction between 1996 and 2008. The open magnetic fields are mostly associated with coronal holes; whereas, the closed magnetic fields that are associated with active regions are more likely to contribute to TSI variations. The long-term variations of the open magnetic field flux is better known, but fortunately the open and closed magnetic fields appear to have similar trends (e.g., Wang et al. 2005). Using the Wang et al. (2005) relationship of total magnetic flux and a reduction of magnetic flux in 2008, the TSI is predicted to be 50 to 150 ppm less in 2008 than in 1996. This larger reduction is based on their model version with varying ephemeral regions, and this model version appears more consistent with the PMOD and ACRIM composite results. Both Tapping et al. (2007) and Steinhilber et al. (2009) also show that lower TSI is expected in 2008 because of reduced magnetic flux during this cycle minimum.

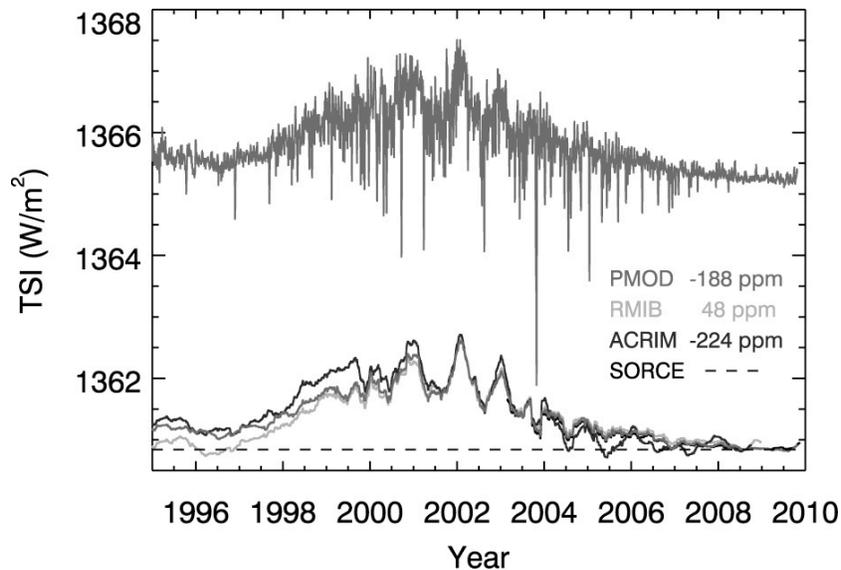

Figure 2. Comparison of TSI time series. The PMOD composite is shown in both daily values on its native scale (upper curve) and with 81-day smoothing adjusted to the SORCE TSI level in 2008 (dashed line). The RMIB and ACRIM composites are only shown as adjusted 81-day smoothing. The changes in the TSI from 1996 to 2008 are listed in the plot; PMOD and ACRIM show about 200 ppm decrease and RMIB indicates a slight increase.

If the TSI and magnetic field relationships are valid (e.g. Wang et al. 2005, Steinhilber et al. 2009), then the lower TSI levels in 2008 from PMOD and ACRIM composites are considered more correct than the higher TSI level from the RMIB composite. However, the conflicting result provides an unique opportunity to challenge our current understanding of instrument calibrations and solar-climate relationships. The TSI is measured with more instruments and with improved accuracy during this minimum than any other previous minima, so additional research has great potential to resolve these differences and to answer the following questions. Is the magnetic field the right proxy to use in reconstructing the secular TSI variations? We expect that solar variability is primarily caused by magnetic activity on the Sun, more so by closed magnetic field than open fields, but our estimate for long-term trends in the solar magnetic field is based primarily on interplanetary magnetic field (IMF) and cosmic rays data that reflect only open magnetic fields. On the instrument front, are the current TSI instruments truly capable of measuring a significant difference between 1996 and 2008? The uncertainty for long-term variations in the TSI record is at best 100 ppm with these TSI composites. These composites do agree very well with the SORCE TIM measurements, which has the lowest degradation rate of only a few ppm/year; therefore, there are expectations that the precision of the composites will improve over time.

While more research into these TSI measurements is expected to better refine these results for 2008, some of the questions posed might not be fully answered until the next generation of TSI instruments have completed measurements over the next cycle. The Glory TIM, Picard SOVAP, and Picard PREMOS instruments are expected to have improved performance in absolute accuracy and long-term stability for the future TSI measurements. Currently, the Picard launch is planned for February 2010, and the Glory launch is expected to be in late 2010.

### 3. Solar Ultraviolet Irradiance Variation

For the first time, there are several concurrent space-based programs with solar irradiance measurements to allow complete spectral coverage during solar cycle minimum conditions. With these measurements, reference spectra from 0.1 to 2400 nm have been assembled for March-April 2008 to represent this cycle minimum irradiance results (Woods et al. 2009, Chamberlin et al. 2009). The primary data sets used are from NASA's Solar Radiation and Climate Experiment (SORCE), Solar EUV Experiment (SEE) aboard the Thermosphere, Ionosphere, Mesosphere, Energetics, and Dynamics (TIMED) satellite, and the underflight SEE calibration rocket payload that also included the prototype EUV Variability Experiment (EVE) developed for the future Solar Dynamics Observatory (SDO) mission.

Unfortunately, there are fewer spectral irradiance measurements during the previous minima. Furthermore, none of the instruments used for the 2008 cycle minimum reference spectrum were operating in 1996 minimum. Therefore, the systematic offsets of a few percent between different instruments make it difficult, if not impossible, to accurately derive how much the long-term spectral irradiance varied from 1996 to 2008. Our previous minimum reference spectrum (Thuillier et al. 2004) is based on ATLAS-3 and rocket measurements that are during low solar activity but not actually during the lowest levels in 1996. So it is no surprise that the comparison of this previous minimum irradiance spectrum is higher than the 2008 measurement. Some attempts have been made to study how the solar ultraviolet (UV) irradiance has varied from 1996 to 2008 using UARS and SORCE measurements, respectively; however, the estimated reductions in 2008 by a few percent in the far ultraviolet (FUV: 120-200 nm) is not significant when considering the uncertainty for these estimates is at best 6% as related to understanding systematic calibration offsets and instrument degradation functions (Marty Snow, private communication, 2009).

The one exception to providing a potentially significant result of solar UV variability between 1996 and 2008 is from the SOHO Solar EUV Monitor (SEM) instrument (Judge et al. 1998). The SEM provides a single-instrument measurement from 1996 to 2009 of the solar EUV irradiance in a broad band from 26 to 34 nm, and the SEM team at the University of Southern California has had several underflight calibration experiments for tracking the instrument degradation. The SOHO SEM result for long-term variability is that the 26 to 34 nm irradiance has decreased from 1996 minimum to 2008 minimum by 15% with an uncertainty of 6% (Leonid Didkovsky, private communication, 2009). While TIMED SEE measurements agree with the SOHO SEM results since TIMED began observations in 2002, there are no other satellite measurements of the solar EUV irradiance during the 1996 minimum. Some supporting evidence for the lower solar EUV irraidance is the reduction of the thermospheric density between 1996 and 2008 based on derivations from satellite drag (Emmert, Picone, & Meier 2008).

Solar EUV images, such as from SOHO EUV Imaging Telescope (EIT), clearly indicate lower radiance from coronal holes (areas of open magnetic field). Thus, one option that might explain lower EUV irradiance is to have more coronal holes. Using CHIANTI spectral models (Dere et al. 1997, Landi et al. 2006) of the quiet sun (QS) and coronal hole (CH) differential emission measures (DEMs), it is estimated that the Sun needs to have 18% more coronal holes in 2008 than in 1996 to explain a 15% reduction of the 26 to 34 nm irradiance. A model estimate for 2008 has 32% QS and 68% CH contributions to match the TIMED SEE measurement (Woods et al. 2005), and the 1996 model uses 50% QS and 50% CH contributions to match a rocket measurement (Woods et al. 1998). The differences between these two models predict the 15% reduction of the SEM 26 to 34 nm irradiance. In addition, the ratio of the 2008 model to 1996 model, as shown in Figure 3 in 1 nm intervals, indicates even more reduction for the XUV 0 to 15 nm irradiance, by as much as 35% for 2008, less reduction for the longer EUV wavelengths, and even a slight increase for the hydrogen emissions. Most of this spectral dependence is expected, based on where most of the brighter coronal contributions emit in the EUV spectrum.

The likelihood of having 18% more coronal holes in 2008 than in 1996 seems low, but the complete analysis of coronal hole evolution from 1996 and 2008 has not been completed. What is most clear is that the polar coronal hole area has decreased by about 20% from 1996 to 2008 (Kirk et al. 2009). But there are several large coronal holes at low-latitudes in 2008; whereas, there were few, if any, low-latitude coronal holes during the 1996 minimum (the more typical case during the previous minima). Some initial analyses of the total coronal hole area suggest a modest, if any, increase of coronal hole area in 2008 as compared to 1996. But the total coronal hole area might not be the most appropriate component for the solar EUV irradiance. For example, the low-latitude coronal holes could have more significant contribution to the solar EUV irradiance than the polar coronal holes that always appear on the limb. Namely, a coronal hole seen on the limb is often obscured in solar EUV images by prominences and spicules, thus the reduction of the solar EUV irradiance is expected to be enhanced when coronal holes are closer to disk center. Therefore, the low-latitude coronal hole area is anticipated to be the more important factor in analyzing the solar EUV irradiance variations. Efforts in analyzing solar EUV images and modeling the solar EUV irradiance are in progress, but results from these efforts are too preliminary for inclusion here.

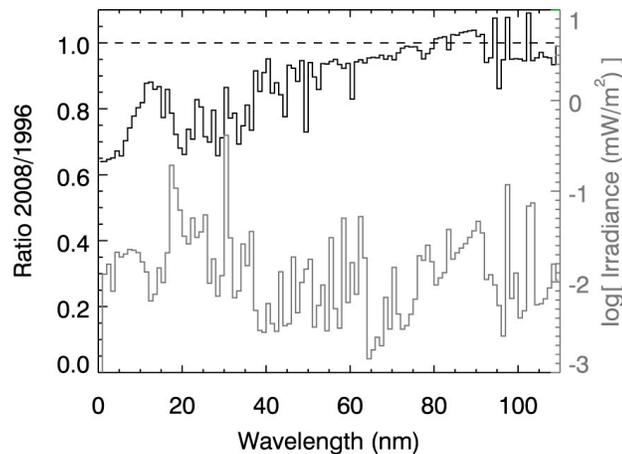

Figure 3. Estimated Variation for the Solar EUV Irradiance Between 1996 and 2008. Increasing the coronal hole area by 18% allows for a 15% decrease in the 26-34 nm band to be consistent with the SOHO SEM measurement from 1996 to 2008. The April 2008 reference spectrum (lower curve) is also presented (Woods et al. 2009).

### 4. Summary

The TSI irradiance appears to be about 200 ppm less in 2008 than in 1996 and is consistent with expectations from reconstructed TSI models based on solar magnetic fields being reduced in 2008. In contrast, there is the RMIB TSI composite that suggests that the TSI is higher in 2008 by about 50 ppm. The uncertainties of these TSI composite time series is at best 100 ppm, thus the long-term variation of the TSI from 1996 to 2008 is not definitely known. There is potential to resolve these differences with additional analysis and modeling efforts.

There are fewer measurements of the solar spectral irradiance variations between 1996 and 2008 than the TSI measurements. The only reasonable result is that the solar 26 to 34 nm irradiance from SOHO SEM appears 15% lower in 2008 than in 1996. This EUV decrease could possibly be explained by the abundance of low-latitude coronal holes during this current cycle minimum, unlike in 1996; however, more image analysis and irradiance modeling is needed to fully address this possibility.

*Acknowledgements* This research is supported by NASA contract NAS5-97045 and NASA grant NAG5-11408 to the University of Colorado. I thank the many SOHO and SORCE scientists who have provided the high quality irradiance data that are discussed in this paper. The PMOD TSI composite is version d41_62_0910 from ftp://ftp.pmodwrc.ch/pub/data/irradiance/composite/DataPlots/. The RMIB TSI composite is from http://remotesensing.oma.be/TSI/data.html. The ACRIM TSI data are from http://www.acrim.com/Data%20Products.htm. The SSN data are from http://www.ngdc.noaa.gov/stp/SOLAR/ftpsunspotnumber.html. The SEM data are from http://www.usc.edu/dept/space_science/sem_data/sem_data.html. The CHIANTI spectral model is a collaborative project involving NRL (USA), RAL (UK), MSSL (UK), the Universities of Florence (Italy) and Cambridge (UK), and George Mason University (USA). Special thanks to Vanessa George for her support in preparing this manuscript.